\documentclass[conference]{IEEEtran}

\usepackage{graphicx}
\usepackage{epsf}

\usepackage{epsfig}

\usepackage{epstopdf}

\newtheorem{definition}{\bf Definition}

\usepackage{ifpdf}

\usepackage{cite}

\ifCLASSINFOpdf
\else
\fi

\usepackage[cmex10]{amsmath}
\usepackage{algorithmic}
\usepackage{array}
\usepackage{mdwmath}
\usepackage{mdwtab}
\usepackage{eqparbox}
\usepackage[tight,footnotesize]{subfigure}
\usepackage{fixltx2e}
\usepackage{stfloats}
\usepackage{url}
\usepackage{caption}
\usepackage{amsfonts}

\usepackage{breakurl}
\usepackage{doi}
\usepackage{float}

\usepackage{amsmath,graphicx}
\usepackage{epsf}

\usepackage{epsfig}

\usepackage{epstopdf}

\begin{document}
%
\title{Context-Aware D2D Peer Selection for Load Distribution in LTE Networks}


\author{\IEEEauthorblockN{Nima Namvar, Niloofar Bahadori, and Fatemeh Afghah}
\IEEEauthorblockA{School of Electrical and Computer Engineering \\
North Carolina A\&T State University, Greensboro, NC 27411, USA\\
Email: \{nnamvar,nbahador\}@aggies.ncat.edu, fafghah@ncat.edu}}
\IEEEtitleabstractindextext{
\begin{abstract}
In this paper we propose a novel context-aware approach for resource allocation in device to device (D2D) communication networks which exploits context information about the users’ velocity and size of their demanded data to decide whether their traffic load can be transferred to D2D tier and which D2D users should be paired. The problem is modeled as a matching game with externalities and a novel algorithm is proposed to solve the game which converges to a stable matching between the D2D users. Simulation results demonstrate the effectiveness of our model in offloading the cellular network’s traffic to the D2D tier.
\end{abstract}

}

\maketitle

\IEEEdisplaynontitleabstractindextext
\IEEEpeerreviewmaketitle

\section{Introduction}
Owning to the ongoing proliferation of user equipments (UEs) such as smartphones and tablets, as well as the recent development of wireless technologies, new mobile multimedia services are becoming available to larger audiences all around the world. The rise of online multimedia services has significantly increased the demand for higher data rate wireless access. This demand is to be met with novel approaches for network design.

Device to device (D2D) communications is emerging as a promising technology to cope with the imminent wireless capacity crunch and increase the spectral efficiency in the next generation of wireless networks \cite{Jeff3GPP}. In the D2D communication, the UE devices are allowed to transmit data signals to each other over a direct link instead of routing through an Evolved Node B (eNB). The main idea is to allow direct D2D
communication over the licensed band and under the control of the cellular system’s operator \cite{D2DZhuHan}. Indeed, by exploiting the natural proximity of communication devices, D2D is capable of yielding multiple performance benefits such as achieving high data rate and low end-to-end delay due to the short range direct communication \cite{Jeff3GPP}.

However, the potential benefits of D2D communication comes at the price of tackling several new technical challenges in the cellular networks.  The introduction of different communication modes (i.e., cellular and D2D) with diverse power, capacity, and range, imposes some kind of heterogeneity to the current well-planned and well-organized cellular systems and leads to many technical challenges such as resource allocation, network modeling, interference mitigation, self-organizing and network economics .

One fundamental issue in D2D communications is how to optimize traffic-offloading from the cellular tier so as to increase the network-wide capacity while also providing consistent and reliable communication for the users \cite{WalidSocial}. However, the mobility of the users can result in intermittent connectivity which degrades the network's quality of service (QoS). Speed and trajectory of the users as well as the size of their transactions are the key factors which determine the expected length and the quality of the wireless connection.

In this paper, the problems of link selection and D2D partner association are considered simultaneously. We exploit the context information about the velocity of the mobile users and the size of their demanded data to calculate the probability of safe communication link between the nearby D2D devices. This probability can assist the network to make a decision on transferring the load from the cellular tier to D2D tier. Furthermore, we formulate the problem of finding optimal D2D pairs as a matching game which assigns a communication link to the neighbouring D2D user pairs. In assigning the communication link, the algorithm seeks to maximize the traffic offloading of the eNB.

\begin{figure}
  \begin{center}
    \includegraphics[width=7cm]{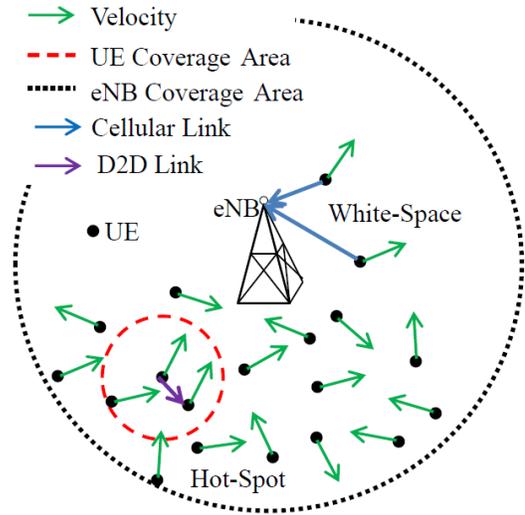}
   \caption{Network Model- D2D enabled UEs move in arbitrary directions in the network seeking reliable D2D links}\vspace{-.5cm}
   \label{fig:NetModel}
  \end{center}
\end{figure}

Simulation results show that the proposed context-aware matching algorithm can significantly offload the cellular network while also maintaining a reliable and consistent communication link for the D2D devices.

The rest of the paper is organized as follows: In section \ref{SEC:Model}, the system model for the proposed scenario is presented. In section \ref{SEC:Matching}, the problem of D2D link assignment is formulated in the framework of matching theory and a novel algorithm for solving the game is proposed. The simulation results are provided in section \ref{SEC:Simulation}. Finally, section \ref{SEC:Conclusion} concludes the paper.

\section{Network Model and Problem Formulation}\label{SEC:Model}
We consider the problem of resource sharing in an orthogonal frequency division multiple access (OFDMA)-based cellular network with an overlaid D2D tier as shown in Figure \ref{fig:NetModel}. Let's $\mathbb{C}=\{c_i\}_{i=1}^{i=N}$ and $\mathbb{D}=\{d_j\}_{j=1}^{j=M}$ represent the set of $N$ orthogonal cellular links being utilized by cellular users and $M$ D2D pairs respectively. The network exploits the frequency division duplex (FDD) technology to manage the uplink and downlink resources. Studies show that under FDD protocol, uplink band is underutilized in comparison to the downlink band specially in congested networks \cite{DopplerModeSelection}. Furthermore, uplink resource sharing in D2D communications only affects the BS and the incurred interference can be mitigated by BS coordination. Indeed, this assumption allows us to bypass the downlink interference to cellular users which is more difficult to handle. We also assume a fully loaded cellular network in which $N$ active cellular users occupy the $N$ orthogonal channels in the cell and there is no spare spectrum.  We note that this assumption can be generalized to the case of non fully loaded networks in which some of the D2D users can be served in the dedicated mode \cite{DopplerModeSelection}.

We assume that both cellular and D2D users have their minimum QoS requirements in terms of SINR. The eNB has all the involved channel state
information available to select the optimal resource sharing strategy between the cellular and the D2D users and to coordinate the transmit power so that the expected throughput is maximized \cite{DopplerResourceSharing}. When a pair of UEs are located in D2D range of one another, they can communicate over a direct D2D link reusing portion of spectrum that is being utilized by cellular transmitters in uplink if the interference caused by the D2D transmitter does not go beyond a predefined threshold $\gamma$ in eNB. It is assumed that an established D2D link cannot be reutilized by another D2D pair. This assumption reduces the overhead signaling for interference management and makes the resource sharing problem easier to handle.

Users travels through the network in arbitrary directions with variable speed in the range of $[0,V_{max}]$. Each UE is equipped with an omnidirectional antenna. For tractability, we assume that the coverage area of each D2D user is a circle with radius $R$. Furthermore, since D2D links are generally established between two nearby UEs, we assume that the D2D links experience an slow Rayleigh fading.

\subsection{Interference Management}

Note that due to orthogonality of resources in OFDMA, there is no intracell interference among the cellular users. Therefore, there are only two types of interference in the proposed model; first, the one of the D2D transmitter to the BS and second, the one of the cellular transmitter to the D2D receiver as depicted in Figure \ref{fig:NetModel}. D2D users are required to keep their interference in the uplink under some threshold $\gamma$. Any possible D2D link is utilized if it can provide a minimum acceptable SINR, $\alpha$,  at the D2D receiver.

The SINR at the base station (BS) is given by:
\begin{equation}\label{SINR-BS}
  SINR_{BS}=\frac{g_{c_iB}P_{c_i}}{N_{0}+g_{d_jB}P_{d_j}}
\end{equation}
where $P_{c_i}$ and $P_{d_j}$ stand for the transmission power of cellular user $c_i$ and D2D transmitter $d_j$ while $g_{c_iB}$ and $g_{d_jB}$ represent their corresponding channel gains to base station (BS). $N_{0}$ is the power of additive white Gaussian noise.

SINR at the BS must be greater than $\gamma$. This threshold determines the maximum possible transmit power of D2D transmitter as follows:
\begin{equation}\label{Max P-D2D}
  P_{d_j}<\frac{1}{g_{d_{j}B}}\left(\frac{g_{c_{i}B}P_{c_i}}{\gamma}-N_{0}\right)
\end{equation}
Note that the D2D pairs use the maximum possible transmit power to optimize their quality of communication over the direct link. Therefore, the maximum achievable SINR at the receiver side of D2D link is given by:
\begin{equation}\label{SINR D2D}
  SINR_{D2D_j}=\frac{g_{d_j}P_{d_j}}{N_{0}+g_{c_id_j}P_{c_i}}
\end{equation}
where $g_{d_j}$ is the channel gain over the direct D2D link for D2D pair $d_j$ and $g_{c_id_j}$ represents the channel gain between the cellular transmitter $c_i$ and D2D receiver $d_j$.

\begin{equation}\label{Max SINR D2D}
  SINR_{D2D_j}^{Max}=\frac{g_{c_iB}g_{d_j}P_{d_j}}{(g_{c_iB}+\gamma g_{c_id_j})N_{0}+\gamma g_{c_id_j}g_{d_jB}P_{d_j}}
\end{equation}

 Since the D2D links are established over relatively short distances, they may reasonably be modeled as a single-slope distance dependent path loss channel, i.e. $g_{d_j}=\frac{K}{D_{j}^\lambda}$, where $K$ is a constant depending on the system's parameters and $\lambda$ is the pathloss exponent and $D_{j}$ is the distance of D2D link. Let's assume that $\alpha$ is the minimum threshold of SINR of D2D links. By substituting the $g_{d_j}$ in the (\ref{Max SINR D2D}), we can calculate the coverage range of D2D transmitter $j$ as:

\begin{equation}\label{R}
  R_{j}=\Big[\frac{Kg_{c_iB}g_{d_j}}{\alpha\left[(g_{c_iB}+\gamma g_{c_id_j})N_{0}+\gamma g_{c_id_j}g_{d_jB}P_{d_j}\right]}\Big]^{\frac{1}{\lambda}}
\end{equation}

\subsection{Consistency of D2D Links}
Given the context information, i.e. the velocity and data size of each UE $i\in \mathcal{N}$, in this section we derive the probability of consistency of each link. By definition, a D2D link is consistent during a communication session if its associated SINR remains above a minimum acceptable threshold during the transaction. Consequently, the probability of link-consistency plays an important role in resource allocation scheme and gives us a measure to guarantee a reliable D2D link for the users.

Assume that two D2D users with velocities $\overrightarrow{V_1}$ and $\overrightarrow{V_2}$ are passing one another with angel $\theta$ as shown in Figure \ref{fig:ThetaModel}. Since $\theta$ can arbitrarily take any value in the interval $(-\frac{\pi}{2}, \frac{\pi}{2})$, it can be modeled as a uniformly distributed random variable with the distribution $f_\theta(\theta)=\frac{1}{\pi}$ over $\theta \in(-\frac{\pi}{2}, \frac{\pi}{2})$. The interaction time, i.e. the duration of time they are in D2D range of one another, is given by:\vspace{-.2cm}
\begin{equation}\label{interaction time}
  T_{ij}=\frac{2R_j\cos(\theta)}{\|\overrightarrow{V_i}-\overrightarrow{V_j}\|}
\end{equation}
Let $S$ denotes the size of transaction being transmitted over the D2D link $l_{ij}$. The link $l_{ij}$ is consistent if and only if $T_{ij}>\frac{S}{\mathcal{C}_{ij}}$ where $\mathcal{C}_{ij}$ is the transmission rate on link $l_{ij}$. Therefore, the probability of a having a consistent link is:

\begin{equation}\label{proability}
  P_{ij}=\int_{-cos^{-1}(\frac{VS}{2R_j\mathcal{C}_{ij}})}^{cos^{-1}(\frac{VS}{2R_j\mathcal{C}_{ij}})}f_\theta(x)dx=\frac{2}{\pi}cos^{-1}(\frac{VS}{2R_j\mathcal{C}_{ij}})
\end{equation}
where $V=\|\overrightarrow{V_i}-\overrightarrow{V_j}\|$ is the relative velocity of the UEs. The consistency probability of each link $l_{ij}$, is a measure of how safe it is for the D2D pair to communicate over this link without interruption. We consider a threshold probability for link consistency, $P_{Th}$, beyond which the link is considered consistent to be exploited by D2D users.

\subsection{Problem Formulation}
 One of the main objectives of D2D communication is to offload the wireless traffic form the cellular network to the ad-hoc D2D tier. However, to assess the performance of a designed network in offloading the traffic from the cellular tire, one needs a measure to quantify the concept of traffic in wireless networks. In this paper, we consider the size of data being circulated among the users as the measure of traffic in the network. Therefore, the goal is to transfer as much wireless traffic as possible from the cellular network to D2D tier without sacrificing the network-wide Qos. Let's assume that $\mu$ is a link-assignment strategy that determines which D2D user can be coupled with another D2D user in its vicinity to form a D2D link. Traffic-offloading depends on the partner-assignment strategy because $\mu$ is the main function that establish the D2D network based on some given conditions and constraints such as QoS requirements and interference. Formally, the link assignment problem can be stated as the following optimization problem:

 \begin{figure}
  \begin{center}
    \includegraphics[width=5.5cm]{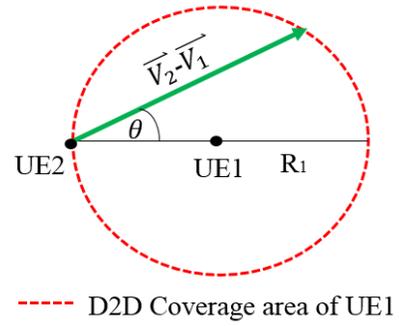}
   \caption{The D2D coverage range and relative motion of the UEs}\vspace{-.5cm}
   \label{fig:ThetaModel}
  \end{center}
\end{figure}

 \begin{equation}\label{Optimization}
 \begin{aligned}
 & \underset{\mu}{\text{maximize}} & & \displaystyle \sum_{i=1}^{N} \displaystyle \sum_{j=1}^{M} \lambda_{ij}(\mu)X_{l_{ij}}\\
 & \text{subject to} & &  \displaystyle \sum_{i} \lambda_{ij}(\mu)\leq 1, \; \lambda_{ij}(\mu)\in\{0,1\}, \; \forall i \in \mathbb{C} \\
 &  & & \displaystyle \sum_{j} \lambda_{ij}(\mu)\leq 1, \; \lambda_{ij}(\mu)\in\{0,1\}, \; \forall j \in \mathbb{D}\\
 &  & & \text{SINR}(BS)\geq\gamma\\
 &  & &\text{SINR}(l_{ij})\geq \alpha\\
 &  & &P_{ij}>P_{Th}\\
 \end{aligned}
\end{equation}

Where $\lambda_{ij}(\mu)$ is an indicator function which is equal to $1$ if the link $l_{ij}$ is selected by $\mu$  and is zero otherwise. $X_{l_{ij}}$ represents the traffic being offloaded to D2D tier if the link $l_{ij}$ is established and is defined as:
\begin{equation}\label{LinkScore}
  X_{l_{ij}}=P_{ij}\mathcal{C}_{ij}T_j
\end{equation}
where $P_{ij}$, $\mathcal{C}_{ij}$, and $T_j$, stand for the consistency probability of link $l_{ij}$, the transmission rate over the link $l_{ij}$, and the interaction time of D2D pair $j$, respectively.

Note that the above optimization problem is subject to context dependent QoS constraints. Namely, the first constraint in (\ref{Optimization}) ensures that each D2D pair are allowed to exploit at most one cellular link. The second constraint ensures that each cellular link can be assigned to at most one D2D pair. The third and fourth conditions capture the minimum SINR requirement for BS and D2D pairs respectively. Finally, the last constraint accounts for the consistency of D2D links.

In terms of complexity, the optimization problem (\ref{Optimization}) is NP hard which depends on the number of cellular links and D2D users in the network. Fortunately, we observe that our proposed link assignment problem can be solved by exploiting tools from matching theory. Specifically, using well-defined utility functions, we introduce a one-to-one matching game between the set of available cellular links and the set of D2D users which converges to optimal link association solution.

\section{Matching Game for D2D Link Assignment}\label{SEC:Matching}
Consider the set $\mathbb{C}=\{c_1,c_2,...,c_N\}$ of all $N$ orthogonal channels and let $\mathbb{D}=\{d_1,d_2,...d_M\}$ be the set of $M$ D2D pairs in the network. The outcome of the links association problem is a \textit{matching} between two sets $\mathbb{C}$ and $\mathbb{D}$ which is defined as follows:

\begin{definition} \hspace{-0.1cm}
A matching $\mathcal{\mu}$ is a function from $\mathbb{C}\cup \mathbb{D}$ to $2^{\mathbb{C}\cup \mathbb{D}}$ such that $\forall c\in \mathbb{C}$ and $\forall d\in \mathbb{D}$: (i) $\mu(c)\in \mathbb{D}\cup \emptyset$ and $|\mu(c)|\leq1$, (ii) $\mu(d)\in 2^{\mathbb{C}}$ and $|\mu(d)|\leq 1$, and (iii) $\mu(c)=d$ if and only if $\mu(d)=c$.
\end{definition}
The potential D2D pairs who are not assigned any link will be served on the cellular mode. Members of $\mathbb{C}$ and $\mathbb{D}$ must have strict, reflexive and transitive preferences over the agents in the opposite set \cite{Roth}. However, we note that the set of channels $\mathbb{C}$ belongs to the network and therefore, by the $\mathbb{C}$'s preference we actually mean the preference of the network over D2D pairs.  Exploiting the context information discussed so far, we introduce the following utility functions to effectively capture the preferences of each set.

\begin{equation}\label{utilityNet}
  U^{Net}=\sum_{i=1}^{N} \displaystyle \sum_{j=1}^{M} \lambda_{ij}(\mu)S_{ij}
\end{equation}
where $S_{ij}$ stands for the size of data being offloaded to D2D tier when channel $i$ is assigned to D2D pair $j$. It is seen that the network prefers to establish D2D links which have a bigger size of data to transfer and by doing so, the cellular network is actually offloading its traffic to D2D tier. The D2D utility function is defined as below.
\begin{equation}\label{utilityD2D}
  U_{j}^{D2D}=P_{ij}(\mu)\mathcal{C}_{ij}(\mu)
\end{equation}
where $P_{ij}$ and $\mathcal{C}_{ij}$ are the probability of link consistency and the transmission rate over the link $l_{ij}$.  (\ref{utilityD2D}) captures the idea that the D2D users prefer the links which are more consistent and have higher transmission rate.

For any D2D pair $j\in \mathbb{D}$ a preference relation $\succ_j$ is defined over the set of links $\mathbb{C}$ such that for any two links $c$ and $c'$ and two matchings $\mu,\mu'\in \mathbb{D}\times \mathbb{C}$, $\mu(j)=c$, $\mu'(j)=c'$:

\begin{equation}\label{PreferenceFunc}
  (c, \mu)\succ_j (c', \mu')\Leftrightarrow U^{D2D}_j(c, \mu)>U^{D2D}_j(c', \mu')
\end{equation}

Similarly, the preference relation $\succ_c, c\in\mathbb{C}$ is defined over the set of D2D pairs $\mathbb{D}$. D2D pairs rank the available links according to their utility function defined in (\ref{utilityD2D}) and the network ranks D2D pairs according to its own utility (\ref{utilityNet}). Our purpose is to match the D2D pairs to the cellular links so that the preferences of both side be satisfied as much as possible; thereby the network-wide efficiency would be optimized.

To solve a matching game, one suitable concept is that of a stable matching.

\begin{definition}
A matching $\mu$  is blocked by the D2D pair $j$ and link $i$ if $\mu(i)\neq j$ and if $i\succ_j \mu(j)$ and $j\succ_i \mu(i)$. A one-to-one matching is \emph{stable} if it is not blocked by any $(i,j)$ pair.
\end{definition}

In the next section, an efficient algorithm for solving the game is presented, which reaches to a stable matching between D2D pairs and cellular links.

\begin{table}[t] \vspace{-0.3cm}
  \scriptsize
  \centering
  \caption{Proposed Algorithm For The D2D Link Assignment Game}
    \begin{tabular}{p{8cm}}
      \hline
\textbf{Input:} context-aware utilities and the preferences of each set \\
\textbf{Output:} Stable matching between the D2D pairs and cellular links \\
\\
\textbf{Initializing}: No link is assigned to D2D users \\
\\
\textbf{Stage I}: \textbf{Preference Lists Composition}
\begin{itemize}
  \item Neighboring D2D users exchange their context information
  \item D2D links with satisfying probability of consistency and rate are stored as the potential D2D pairs
\end{itemize} \\
\textbf{Stage II}: \textbf{Matching Evaluation} \\
\hspace*{1em}\textbf{while:} $\mu^{(n+1)}\neq \mu^{(n)}$
\begin{itemize}
  \item Update the utilities based on the current matching $\mu$
  \item Construct the preference lists using preference relations
  \item Each D2D user $n$ applies for a link connecting it to its most preferred D2D partner
  \item The network keeps the most preferred link requests of each D2D user and rejects the rest
\end{itemize}
\hspace*{2em} \textbf{Repeat}\\
\hspace*{4em} $\bullet$ Each non-paired D2D user applies to get a link which connects it\\
\hspace*{5em} to its next most preferred partner \\
\hspace*{4em} $\bullet$ The network updates its waiting list considering the new applications \\
\hspace*{5em} and keeps the best applications \\
\hspace*{2em} \textbf{Until:} All the D2D users are either assigned a partner along with a link \\
\hspace*{5em} or their requests are all rejected \\
\hspace*{1em}\textbf{end} \\
   \hline
    \end{tabular}\label{Table:Algorithm}\vspace{-0.6cm}

\end{table}

\subsection{Proposed Algorithm}\vspace{-0.1cm}
The deferred acceptance algorithm, introduced in \cite{Roth}, is a well-known approach to solving the standard matching games. However, in our game, the preferences of agents as shown in (\ref{utilityD2D}) and (\ref{utilityNet}), depend on externalities through the entire matching, unlike classical matching problems in which preferences are static and independent of the matching. Therefore, the classical approaches such as the deferred acceptance cannot be used here because of the presence of externalities \cite{ExternalBando}. To solve the formulated game, we propose a novel algorithm shown in Table \ref{Table:Algorithm}. Suppose that initially no link is assigned to D2D users. Each D2D user exchanges its context information with the neighboring D2D users hopping to establish a D2D link. Given the list of available partners, each D2D user sends a request to the network to get a link which connects it with its most preferred partner. On the other hand, the network ranks all the applications concerning each user and only keeps the one which help it to offload more traffic to D2D tier and rejects the rest. Upon ranking the acceptable requests, the network feeds back the awaiting users with its own context information including the rate of the potential links for each user.

Each user makes a ranking list of the available partners along with their corresponding link condition (i.e. rate and consistency). The users who have been rejected in the former phase, would apply to get a link connecting them with their next favorite partner and the eNB modifies its waiting list accordingly. This procedure continues until all the users assigned to a waiting list.

However, since the preferences depend on the current matching $\mu$, an iterative approach should be employed. In each step, the utilities would be updated based on the current matching. Once the utilities are updated, the preference lists would be updated accordingly as well. Therefore, in each iteration, a new temporal matching arises and based on this matching, the interdependent utilities are updated as well. The algorithm initiates the next iteration based on the modified preferences. The iterations run on until two subsequent temporal matchings are the same and algorithm converges.

The proposed algorithm will lead to a stable matching when it converges. Indeed, the deferred acceptance in stage II would not converge if the matching is not stable \cite{Roth}. Hence, by contradiction, whenever the algorithm converges, the matching would be stable.

\section{Simulation Results}\label{SEC:Simulation}
For simulations, we consider a normalized circular cell (i.e. radius $1$) and discuss the offloading feature of D2D tier according to the proposed algorithm. The OFDMA network is capable of serving $100$ cellular users using orthogonal resources. We assume that all users have D2D capability and can switch between D2D and cellular mode depending on the perceived QoS. For simplicity, we consider the single slope path loss channel model with path loss exponent $2$. We also restrict the distance between D2D users to enable a clearer presentation of the results.

\begin{figure}
  \begin{center}
    \includegraphics[width=9cm]{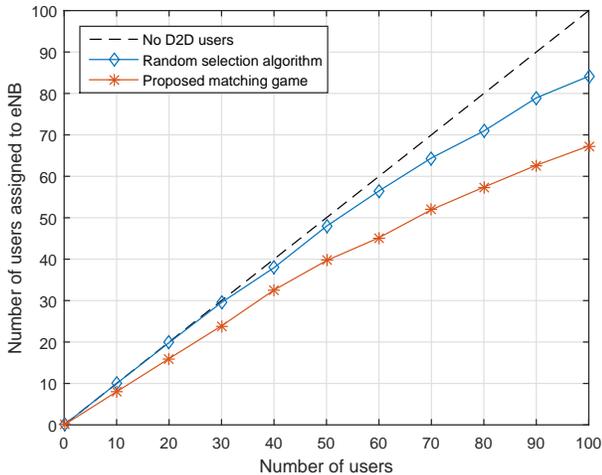}
   \caption{Average number of users assigned to cellular tier}\vspace{-.2cm}
   \label{sim1}
  \end{center}
\end{figure}

Figure \ref{sim1} shows the average number of users assigned to the cellular tier. It shows that as the number of the users increases and the network becomes congested, more users can be offloaded from the cellular network to the D2D tier. It can be seen that the proposed matching game yields considerable performance gains over the random assignment strategy which randomly select D2D pairs and can offload the cellular traffic up to $32.7\%$ to the D2D tier  for a network size of $N=100$ users.

\begin{figure}
  \begin{center}
    \includegraphics[width=9cm]{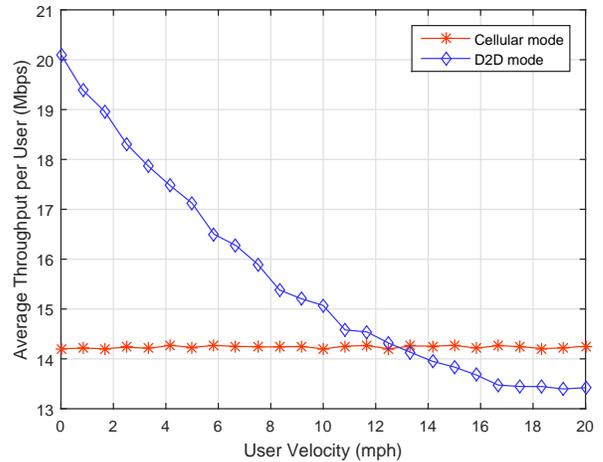}
   \caption{Average throughput per user as a function of users' velocity}\vspace{-.2cm}
   \label{sim2}
  \end{center}
\end{figure}

Figure \ref{sim2} shows average throughput per user as a function of users' velocity. It shows that as the velocity of users increases, the average throughput of D2D users decreases due to the lower probability of finding a consistent link. We note that increasing the velocity beyond some point, i.e. $13$ mph, results in lower throughput in D2D mode compared to cellular mode. This observation is in agreement with the fact that D2D communication is suitable for congested areas in the cellular network in which the users have low to medium mobility.

\section{Conclusion}\label{SEC:Conclusion}
In this paper, we have proposed a novel D2D peer selection approach which exploits the context information about the users' velocity and size of their demanded data to assign a link to the D2D pairs. We derived a closed-form formula for the probability of link-consistency which has a significant impact on the QoS of D2D links. We formulated the problem as a matching game and proposed a novel distributed algorithm to reach to a stable matching between the set of D2D user and available links. Simulation results show that the proposed algorithm yields considerable gains in terms of traffic offloading from the cellular network and the average user throughput compared to that of the random partner assignment scenario for the D2D users.

\section{Acknowledgement}
The first and second authors would like to acknowledge the support from Air Force Research Laboratory and OSD for sponsoring this research under agreement number FA8750-15-2-0116. The U.S. Government is authorized to reproduce and distribute reprints for Governmental purposes notwithstanding any copyright notation thereon. The views and conclusions contained herein are those of the authors and should not be interpreted as necessarily representing the official policies or endorsements, either expressed or implied, of Air Force Research Laboratory, OSD, or the U.S. Government.

\bibliography{myref}
\bibliographystyle{IEEEtran}

\end{document}